# Quantum Dot Solar cells

Husien Salama[1]

*husien.salama@uconn.edu*

**Abstract**

There remains wide interest in solar cells being made using inexpensive materials and simple device manufacturing techniques to harvest ever-increasing amounts of energy. New semiconductor materials and new quantum nanostructures are exploited to fabricate high-efficiency next-generation solar cells. Quantum dots have offered an attractive option for photovoltaics. A single photon absorbed by a quantum dot produces more than one bound electron-hole pair, or exciton, thereby doubling normal conversion efficiency numbers seen in single-junction silicon cells. One potential is to use the solar cell configuration which incorporates.

Quantum Dots Super Lattice (QDSL). QDSL provides a mechanism for the enhancement of solar sales

due to their higher band gap and absorption coefficient when compared to their bulk material counterpart. The mini bands in the conduction, as well as the valence band of a QDSL, play an important role in solar cells because the photogenerated carriers are collected via transport using mini bands. In recent years, a lot of research has been done on crystal growth, structural, electrical, and optical properties of thin films and nanostructures as well as fabrication processes and characterization of photovoltaic devices. In this paper, all the recent developments in future generation quantum dot solar cells like a tandem, intermediate band, and solution-processed band alignment engineering from several research works will be presented.

1. Introduction

**Nanomaterials** of a size comparable to the exciton Bohr radius behave as a quantum dot (QD) due to the three-dimensional quantum confinement of carriers and exhibit a quantum-size effect that provides the opportunity to engineer the bandgap energy (Eg) by adjusting the size of the QD [1]. Minibands can appear in closely packed and well-aligned QD superlattices due to the overlap of confined energy levels. The absorption edge can be tuned in a wide range of photon energies due to the quantum size effect, so it is possible to apply nanocrystal materials to all tandem solar cells, which have the possibility to overcome the Shockley-Queasier limit [2]. Moreover, it has been found that the weak absorption in the bulk material is significantly enhanced in nanocrystals, especially in the small dot size, due to the quantum confinement-induced mixing of Γ-character into the X-like conduction band states. Bulk Silicon (Si) has a band gap of 1.11 eV while Si QD has an effective band gap of 1.24 eV at 4nm with an absorption coefficient of roughly 3 times greater. Ge QDSL structures also have a higher effective band gap of 0.85 eV compared to 0.67 eV for bulk properties. By tailoring the electronic band structure of highly mismatched alloys, researchers have shown clear evidence of the existence of three electronically isolated energy bands, bringing the intermediate-band solar cell one step closer to realization. The Intermediate band solar cell with multi-stacked InGaAs/GaNAs QDs grown using strain compensation techniques has shown improved efficiency in the evolution of solar cells [3]. Solution processing is a promising route for the realization of low-cost, large-area, flexible and lightweight photovoltaic devices with short energy payback time and high specific power. By engineering the band alignment of the quantum dot layers using different ligand treatments, a certified efficiency of 8% has been reached [9].

2. Tandem cells

One of the promising methods to enhance the efficiency of solar cells is to use a stack of solar cells, in which each cell has a band gap that is optimized for the absorption of a certain spectral region the importance of multijunction solar cells is that both spectrum splitting and photon selectivity are automatically achieved by the stacking arrangement. To achieve the highest efficiency from the overall tandem device, the power from each cell in the stack must be optimized.

This is done by choosing appropriate bandgaps, thicknesses, junction depths, and doping characteristics, such that the incident solar spectrum is split between the cells most effectively [5].

The phosphorous doping in n-type Si QDs superlattice was realized by P2O5 co-sputtering during the deposition of silicon-rich oxide (SRO, Si, and SiO2 co-sputtering), which forms Si QDs upon high-temperature post-annealing [10].

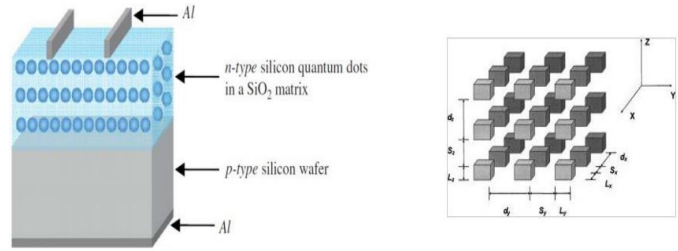

Fig 1(a): Schematic diagram of (n-types) Si QDs and (p-type) c-Si heterojunction solar cell (b) Si QDSL array [5]

For QDs which are close to each other such that there is a significant tunnel probability, a true miniband will form and result in an increase in the effective band gap of the tailored superlattice material. The conceptual design of a complete device is shown in Figure 2.

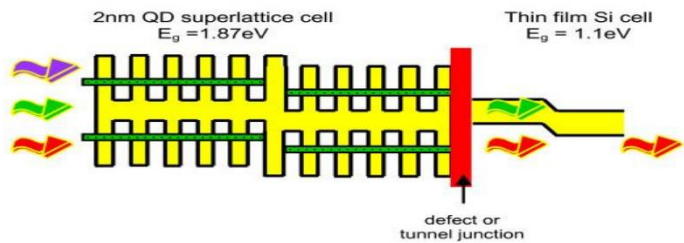

Fig 2: Si QDs in an amorphous dielectric matrix showing a defect tunnel junction connection to a thin-film Si cell [6]

## 2.1 Development in Fabrication techniques:

Si QD materials grown by sputtering alternating Si-rich oxide and stoichiometric SiO2 layers, with a subsequent, anneal to precipitate a layered growth of QDs, have been used earlier [6]. Absorption data on these show a strong absorption region with absorption dependent on Si content and a significant weak absorption tail attributed to sub-oxide regions surrounding QDs due to incomplete diffusion to the growing QDs. Dramatic decreases in resistivity have been demonstrated on the incorporation of P or B dopants and rectifying p-n junctions have been fabricated which give a VOC up to 490 mV. However, the mechanism of doping in these materials can neither be direct doping of the QDs nor modulation doping of a stoichiometric SiO2 matrix. A modified modulation doping mechanism is proposed in which P or B atoms dope the sub-oxide region shown to surround QDs, which because of its reduced band gap has dopant levels shallow enough to be ionized and hence provide free carriers to the QDs [11].

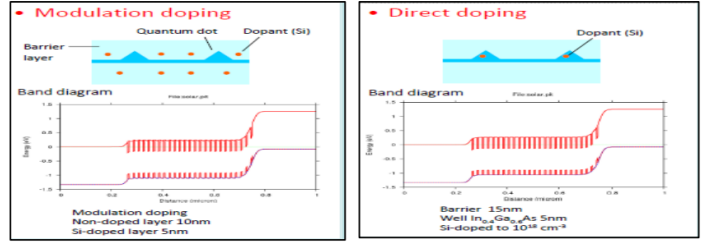

Fig 3: Comparison between modulation and direct doping [4]

An advanced top-down process, bio-template (Li-Dps), and damage-free neutral-beam techniques to successfully fabricate a highly ordered and dense 2D array of 6.4-nm-diameter SiNDs with a SiC interlayer have been used [6]. The Si-ND array with SiC interlayer demonstrated high optical absorption and high electrical transport due to the formation of a miniband. Furthermore, the paper reports for the first time a high-efficiency Si quantum dot solar cell with 12.6% energy conversion.

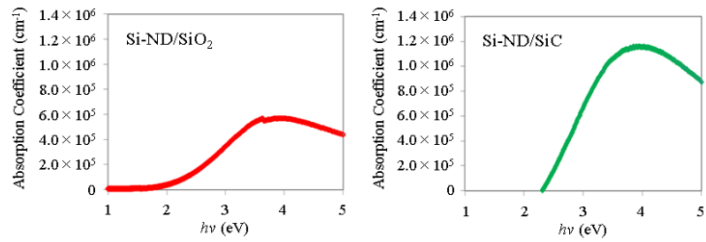

Fig 4: Absorption coefficient of 4-nm-thick (6.4-nm diameter) Si-ND with 3-nm-thick SiO2 and 3-nm-thick SiC structures. [1].

## 3. Intermediate band Solar cells

The intermediate-band solar cell (IBSC) is one of the most promising candidates for the next generation of photovoltaic devices. These solar cells involve a two-step transition: A photon with energy higher than the bandgap between the valence and the intermediate bands is absorbed to generate an electron. Then, another photon with energy higher than the bandgap between the intermediate and the conduction bands pumps the previous electron to the conduction band. By providing an additional intermediate band between the valence and the conduction bands, photon absorption has been made more efficient [3]. QD solar cell is one of the methods to implement IBSCs. The quantum dots are embedded in a matrix to form an intermediate band, with its energy level depending on the QD size.

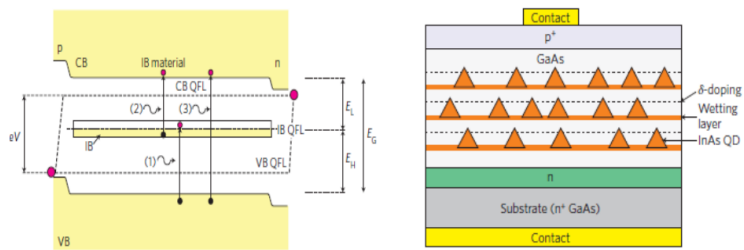

Fig 5: IBSC energy band diagram and device structure. [3]

## 3.1 Challenges

The primary issue in all reports of QD solar cells is that Voc decreases with respect to reference cells without QDs, even though the short-circuit current density shows a small increase in devices with QDs. Voc reduction may be related to the enhanced generation of the saturation current, which is determined by various mechanisms, such as the recombination current or diffusion current. In either case, Voc decreases with increasing temperature, as determined by the bandgap energy of the host semiconductor, which is denoted by Eg, and the dark current Characteristics [7].

$$\mathbf{Voc\ (T) = Eg\ /\ q - CT} \tag{1}$$

where C is the temperature coefficient, which reflects the dark current characteristics in the solar cells.

**Table 1: InAs/GaNAs Characteristics**

|  | $V_{oc}$ (V) | $I_{sc}$ (mA/cm²) | FF (%) | η (%) | Diode factor |
|---|---|---|---|---|---|
| 20 stacks | 0.72 | 21.04 | 70.0 | 10.63 | 1.65 |
| 30 stacks | 0.67 | 22.33 | 70.76 | 10.59 | 1.67 |
| 50 stacks | 0.68 | 26.36 | 70.24 | 12.44 | 1.59 |

The multi-stacked QDSC are fabricated using the strain compensation technique. Unfortunately, increasing the number of QD layers causes an accumulation of strain that seriously damages the cell's structure. It is strain arising from the lattice mismatch between GaAs (0.567 nm) and InAs (0.604 nm) that promotes the Stranski–Krastanov mode of growth. GaNAs layers were added to achieve strain compensation, but this material has a smaller bandgap than GaAs [3]. Thus, although the cell current increased, its voltage was limited to the GaNAs bandgap as seen in Table1.

## 3.2 Alternative Techniques

InGaAs QDs are alternative candidates for multi-stacked structures because of their small lattice mismatch with GaAs substrates. In earlier works, the fabrication of a highly stacked and well-aligned In0.4Ga0.6As QD structure of over 100 layers without employing a strain balancing technique has been reported [8]. The structure was fabricated by the intermittent deposition of In0.4Ga0.6As layers and by an As2 source, and as a result, there was no degradation in crystal quality. Moreover, miniband formation in 20-stack In0.4Ga0.6As QD superlattices with an interspacing of 3.5nm was achieved. Miniband formation was confirmed by the excitation power dependence observed in photoluminescence (PL) measurements.

A conversion efficiency of 12.6% is obtained for a QD solar cell with an interdot spacing of 35nm, which is larger than that of Table I parameters. The short-circuit current density increases from Jsc 17.8 to 18.6 and 19.0 mA/cm2 as the interior spacing decreases. All the Voc values for the In0.4Ga0.6As QD solar cells are above 0.77V for the various interdot spacings higher than InAs/GaNAs parameters. These results indicate that highly stacked In0.4Ga0.6As QDs are suitable for solar cell devices with a mini band, which allows the carriers to tunnel through the barrier layer and collect at the electrode.

## 4. Colloidal QD solar cells

Solution processing is a promising route for the realization of low-cost, large-area, flexible and lightweight photovoltaic devices with short energy payback time and high specific power [9]. However, solar cells based on solution-processed organic, inorganic, and hybrid materials reported thus far generally suffer from poor air stability, require an inert-atmosphere processing environment, or necessitate high-temperature processing, all of which increase manufacturing complexities and costs. The development of room-temperature solution-processed ZnO/PbS quantum dot solar cells has been successful. The novel solar cell employs layers of QDs treated with different ligands for different functions by tuning their relative band alignment - a layer of inorganic ligand-passivated QDs serves as the main light-absorbing layer and a layer of organic-ligand-passivated QDs serves as an electron-blocking/hole-extraction layer [9].

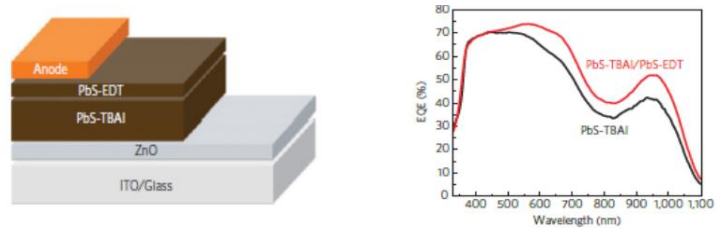

Fig 6: Device architecture and External Quantum efficiency spectra [9]

Oleic-acid-capped PbS QDs with the first exciton absorption peak at λ= 901nm in solution are used to fabricate the thin films. Tetrabutylammonium iodide (TBAI) and 1,2-ethanedithiol (EDT) are used as the inorganic and organic ligands for solid-state ligand exchange. After solid-state ligand exchange, the first exciton absorption peak shifts to λ=935 nm, which corresponds to an optical bandgap Eg= 1.33 eV. The band alignment demonstrates the role of the PbS-EDT layer as an electron-blocking/hole-extraction layer between the PbS-TBAI layer and the anode, which leads to an improved photocurrent collection efficiency and enhanced device performance in the PbS-TBAI/PbS-EDT devices.

The device stability is found to depend to a greater extent on the interface and band alignment between the QDs and anodes than on the bulk QD layer itself. By engineering the band alignment of the quantum dot layers using different ligand treatments, a certified efficiency of 8.55% has been reached [9].

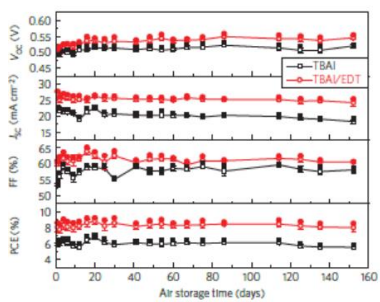

Fig 7: Long-term stability assessment data [9].

Table 3: Merits and limitations of PbS-TBAI solar cell

| Merits | Limitation |
| --- | --- |
| 1. Low cost, simple room temperature fabrication process. | 1. Electrode deposition cannot be done in room temperature. |
| 2. Good air-storage stability. | 2. Relatively low $V_{oc}$ |

By elucidating the origin of the low VOC, optimizing combinations of ligands and QD sizes, and further improving surface passivation via solution-phase treatments will result in continued efficiency improvements. A greater understanding of the QD optoelectronic properties and further progress in materials development could lead to a generation of air-stable, solution-processable QD-based solar cells.

## 5. Conclusion

Solar-cell technology has advanced rapidly, as hundreds of groups around the world pursue more than two dozen approaches using different materials, technologies, and approaches to improve efficiency and reduce costs. To obtain high performance, extensive research work is needed, and it will take some time to realize high-efficiency solar cells containing third-generation concepts. Simultaneously fulfilling the goals of high efficiency, low-temperature fabrication conditions, and good atmospheric stability remains a major technical challenge. Several new works represent a significant leap in overcoming process limitations, increasing the current flow in the cells and thus boosting their overall efficiency in converting sunlight into electricity. Silicon had six decades to get where it is today, and even silicon hasn't reached the theoretical limit yet. So, the newer technology cannot beat an incumbent in just a few years of development. But it can be clearly seen that quantum dots have a very good potential in solar cells. There is still a long way to go before quantum-dot solar cells are commercially viable, but several milestones have been achieved in recent developments. A green revolution in the renewable energy domain using low-cost QD solar cells is not far from being realized [12-13].

# 6. References

1. Seiji Samukawa, "Quantum Dot Superlattice Structure for High Efficiency Solar Cells Utilizing a Bio-template and Damage-free Neutral Beam Etching", (ICSICT) IEEE 2012.

2. Shigeru Yamada, Yasuyoshi Kurokawa, Shinsuke Miyajima and Makoto Konagai, "Silicon quantum dot superlattice solar cell structure including silicon nanocrystals in a photogeneration layer", nanoscale research letters 2014.

3. Antonio Luque, Antonio Martí & Colin Stanley, "Understanding intermediate-band solar cells", Nature Photonics 6, 146-152(2012).

4. Yoshitaka Okada, "Quantum Dot Superlattice Intermediate band cells",Nanophotonics conference, 19, 2010.

5. Foozieh Sohrabi, Arash Nikniazi and Hossein Movla, Optimization of Third Generation Nanostructured Silicon-Based Solar Cells, http://dx.doi.org/10.5772/51616

6. Gavin Conibeer, Ivan Perez-Wurfl,Xiaojing Hao,Dawei Di,and Dong Lin, "Si solid-state quantum dot-based materials for tandem solar cells", Nanoletters 2012.

7. Takeshi Tayagaki, Yusuke Hoshi and Noritaka Usami, "Investigation of the open-circuit voltage in solar cells doped with quantum dots", Nature 2013

8. T. Sugaya, O.Numakami, S.Furue, H.Komaki, T.Amano, K.Matsubara, Y.Okano, S.Niki "Tunnel current through a miniband in InGaAs quantum dot superlattice solar cells" Solar Energy Materials & Solar Cells 95, 2920-2923 (2011)

9. Chia-Hao M. Chuang, Patrick R. Brown, Vladimir Bulovic and Moungi G. Bawendi, "Improved performance and stability in quantum dot solar cells through band alignment engineering", Nature materials May 2014

10. Salama, H., B. Saman, R. Gudlavalleti, R. Mays, E. Heller, J. Chandy, and F. Jain. "Compact 1-Bit Full Adder and 2-Bit SRAMs Using n-SWS-FETs." International Journal of High-Speed Electronics and Systems 29, no. 01n04 (2020): 2040013.

11. Salama, H., B. Saman, R. H. Gudlavalleti, P. Y. Chan, R. Mays, B. Khan, E. Heller, J. Chandy, and F. Jain. "Simulation of Stacked Quantum Dot Channels SWS-FET Using Multi-FET ABM Modeling." International Journal of High-Speed Electronics and Systems 28, no. 03n04 (2019): 1940025.

12. Gudlavalleti, R. H., B. Saman, R. Mays, H. Salama, Evan Heller, J. Chandy, and F. Jain. "A novel addressing circuit for SWS-FET based multivalued dynamic random-access memory array." International Journal of High-Speed Electronics and Systems 29, no. 01n04 (2020): 2040009.

13. Salama, Husien, Bandar Saman, Evan Heller, Raja Hari Gudlavalleti, Roman Mays, and Faquir Jain. "Twin Drain Quantum Well/Quantum Dot Channel Spatial Wave-Function Switched (SWS) FETs for Multi-Valued Logic and Compact DRAMs." International Journal of High-Speed Electronics and Systems 27, no. 03n04 (2018): 1840024.